\documentclass[aps,pre,showpacs,showkeys,twocolumn,superscriptaddress,
              floats,floatfix,preprintnumbers,longbibliography]{revtex4-1}
\usepackage{amssymb,amsmath}
\usepackage{bm,url,graphics,subfigure,epsfig,rotating,float}
\usepackage{ulem}
\usepackage{color,soul}
\def \BB {\mathcal{B}}
\def \tplus {t^{+}}
\def \tminus {t^{-}}
\def \Qplus {Q^{+}}
\def \Qminus {Q^{-}}
\def \eps {\varepsilon}
\def \Eflow {E_{\rm flow}}

\def \kmax {k_{\rm max}}
\def \tgain {t_{\rm gain}}
\def \tloss {t_{\rm loss}}

\def \rhof {\rho_{\rm f}}
\def \rhop {\rho_{\rm p}}

\def \uu  {{\bm u}}
\def \vv  {{\bm v}}
\def \ff  {{\bm f}}

\def \teta {t_{\rm \eta}}

\def  \xx  {{\bm x}}
\def  \XX  {{\bm X}}
\def \Xdot {\dot{{\bm X}}}
\def \vdot {\dot{{\bm v}}}

\def \taup {\tau_{\rm p}}

\def \grad {{\bm \nabla}}

\def \lap {\nabla^2}

\newcommand{\bra}[1]{\langle #1\rangle}
\def \Ir  {\mbox{Ir}}
\def \Rey {\mbox{Re}}

\def \St  {\mbox{St}}
\def \StT  {\mbox{St}_{\rm T}}

\def \Teddy {T_{\rm eddy}}

\def \urms  {u_{\rm rms}}

\newcommand{\Eq}[1]{Eq.~(\ref{#1})}
\newcommand{\Fig}[1]{Fig.~(\ref{#1})}
\newcommand{\subfig}[2]{Fig.~(\ref{#1}#2)}

\DeclareMathOperator{\Tr}{Tr}
\DeclareMathOperator{\Det}{Det}
\newcommand{\bfig}{\begin{figure}}
\newcommand{\efig}{\end{figure}}
\newcommand{\bc}{\begin{center}}
\newcommand{\ec}{\end{center}}
\newcommand{\bea}{\begin{eqnarray}}
\newcommand{\eea}{\end{eqnarray}}

\begin{document}
\title{Heavy inertial particles in turbulent flows gain energy slowly but lose 
it rapidly}
\author{Akshay Bhatnagar}
\email{akshayphy@gmail.com}
\affiliation{Centre for Condensed Matter Theory, Department of Physics, Indian Institute of Science, Bangalore 560012, India.}
\affiliation{Nordita, KTH Royal Institute of Technology and
Stockholm University, Roslagstullsbacken 23, 10691 Stockholm, Sweden}
\author{Anupam Gupta}
\email{anupam1509@gmail.com}
\affiliation{Mechanical Science and Engineering, University of Illinois, 
1206 W. Green Street, Urbana, IL 61801, USA}
\author{Dhrubaditya Mitra}
\email{dhruba.mitra@gmail.com}
\affiliation{Nordita, KTH Royal Institute of Technology and
Stockholm University, Roslagstullsbacken 23, 10691 Stockholm, Sweden}
\author{Rahul Pandit}
\email{rahul@iisc.ac.in}
\affiliation{Centre for Condensed Matter Theory, Department of Physics, Indian Institute
of Science, Bangalore 560012, India.}
\preprint{NORDITA 2017-117}
\begin{abstract}
We present an extensive numerical study of the time irreversibility of the
dynamics of heavy inertial particles in three-dimensional, statistically
homogeneous and isotropic turbulent flows.  We show that the probability density
function (PDF) of the increment, $W(\tau)$, of a particle's energy over a
time-scale $\tau$ is non-Gaussian, and skewed towards negative values. This
implies that, on average, particles gain energy over a period of time that is
longer than the duration over which they lose energy. We call this \textit{slow
gain} and \textit{fast loss}.  We find that the third moment of  $W(\tau)$
scales as $\tau^3$,  for small values of $\tau$.  We show that the  PDF of
power-input $p$ is negatively skewed too; we use this skewness $\Ir$ as a
measure of the time-irreversibility and we demonstrate that it increases
sharply with the Stokes number $\St$, for small $\St$; this increase slows down
at $\St \simeq 1$. Furthermore, we obtain the PDFs of $\tplus$ and $\tminus$,
the times over which $p$ has, respectively, positive or negative signs, i.e.,
the particle gains or loses energy. We obtain from these PDFs a direct and
natural quantification of the the slow-gain and fast-loss of the particles,
because these PDFs possess exponential tails, whence we infer the
characteristic loss and gain times $\tloss$ and $\tgain$, respectively; and we
obtain $\tloss < \tgain$, for all the cases we have considered.  Finally, we
show that the slow-gain in energy of the particles is equally likely in
vortical or strain-dominated regions of the flow; in contrast, the fast-loss of
energy occurs with greater probability in the latter than in the former.
\end{abstract}
\maketitle
\section{Introduction}
\label{chap4:intro}

Heavy inertial particles (or heavy particles) advected by turbulent flows are
found in many natural phenomena and industrial processes; examples include dust
particles in a storm~\cite{kok2012physics}, water droplets in a turbulent
cloud~\cite{Pruppacher2010microphysics}, pollutant dispersions, the formation
of planetesimals~\cite{Arm10}, and turbulent mixing in chemical
reactions~\cite{sienfeld1986atmospheric,csanady1976turbulent,
bracco1999particle,pinsky1997turbulence,rothschild1988small}.  These heavy
particles cannot be modeled as tracers because of their finite size and
inertia. Many experimental, numerical, and theoretical studies have been
carried out to understand the statistics of these particles in turbulent flows,
\cite[see, e.g.,][for
reviews]{toschi2009lagrangian,gus+meh16,pumir2016collisional}.  Such a system
of heavy particles also displays many intriguing features that are of interest
in nonequilibrium statistical mechanics.  

Some recent studies have investigated the time irreversibility of fluid
turbulence by using the statistics of Lagrangian-tracer
particles~\cite{xu2013power,xu2014flight,
leveque2014introduction,grafke2015time,cencini2017time}.  Fully-developed
Navier--Stokes turbulence occurs in the limit of infinite Reynolds number or
zero viscosity.  The rate of energy dissipation $\eps$ does not go to zero, but
it remains constant even at the highest values of the Reynolds numbers $\Rey$
that have been obtained in experiments and numerical simulations. The
hypothesis $\eps > 0$ as $\Rey \to \infty$, which lies at the core of the
Kolmogorov theory (K41) of turbulence, is known as the zeroth law of
turbulence~\cite{Fri96}. Fully developed forced turbulence is a nonequilibrium,
statistically stationary state, which displays a constant average flux of
energy from large to small length scales, where it is dissipated by viscosity.
Hence, obviously, such turbulence is irreversible in time.  However, this is
not immediately obvious to our eyes, if we look at movies of the advection of
Lagrangian tracers.  By following the evolution of the kinetic energy of a
single tracer particle, Ref.~\cite{xu2014flight} shows that, on average, these
tracers decelerate faster than they accelerate.  This phenomenon of
\textit{slow-gain} and \textit{fast-loss} of energy has been suggested to be
the signature of irreversible, turbulent dynamics, in the trajectory of a
single Lagrangian tracer;  and it has been quantified, indirectly, in
Refs.~\cite{xu2013power,xu2014flight} by the negative third moment of the
probability density function (PDF) of the particle's energy increments, and the
negative skewness of the PDF of the power input $p$ to the particles by the
flow. This observation suggests the violation  of the principle of detailed
balance in turbulent flows. 

It is straightforward to understand this slow-gain and fast-loss phenomenon
qualitatively via the K41 phenomenology of turbulence: The turbulent cascade in
the inertial range conserves energy. The energy is injected into the fluid at
the large, \textit{integral length scale} and dissipated significantly at the
small length scales that lie below the Kolmogorov dissipation scale.  The
eddies at the largest length scales evolve most slowly; and those at the
smallest length scales are the fastest; hence, the dynamics of a single tracer
particle shows the \textit{slow-gain} and \textit{fast-loss} features described
above; and the resulting irreversibility is, therefore, related to the
aforementioned separation of time-scales in turbulent flows. 

We extend these ideas to heavy particles in turbulent flows by carrying out an
extensive numerical study of the time irreversibility of the dynamics of heavy
inertial particles in three-dimensional (3D), statistically homogeneous and
isotropic turbulent flows. In addition to being advected by the
time-irreversible turbulent flow, heavy particles experience a drag force that
introduces an additional source of dissipation.  Nevertheless, it is still
impossible to distinguish visually between forward-in-time and backward-in-time
trajectories of individual particles.  We illustrate this in videos
V1~\cite{videoV1} and V2~\cite{videoV2} for representative heavy-particle 
trajectories in statistically stationary turbulent flows that are homogeneous 
and isotropic;
video V1 runs forward in time and V2 runs backwards.
However, merely by looking at the two videos it is not possible to tell which
one is which.  Following Refs.~\cite{xu2013power,xu2014flight}, which consider
Lagrangian tracers, we first characterize the irreversibility of the
trajectories of heavy particles by the following two quantities: (a) The energy
difference of a particle across a time scale $\tau$,
\begin{equation}
W(\tau) = E(t+\tau) - E(t) \/,
\label{eq:W}
\end{equation}
where $E(t)$ is the energy per unit mass of the particle at time $t$; and (b)
the skewness of the PDF of the power input $p$ to the particle by the flow.  We
show that the probability density function (PDF) of the increment, $W(\tau)$, of
a particle's energy over a time-scale $\tau$ is non-Gaussian, and skewed
towards negative values. This implies that, on average, particles gain energy
over a period of time that is longer than the duration over which they lose
energy. We call this \textit{slow gain} and \textit{fast loss}. We find that
the third moment of the PDF of $W(\tau)$ is negative and scales as $\tau^3$,
for small values of $\tau$.  Next, we calculate the PDFs of times over which
the power $p$ retains the same sign. In particular, we show that the  PDF of
$p$ is negatively skewed; we use this skewness $\Ir$ as a measure of the
time-irreversibility and and we demonstrate that it increases sharply with the
particle Stokes number $\St$ (see below), for small $\St$; this increase slows
down at $\St \simeq 1$. Furthermore, we obtain the PDFs of $\tplus$ and
$\tminus$, the times over which $p$ has, respectively, positive or negative
signs, i.e., the particle gains or loses energy. From these PDFs we obtain a
direct and natural quantification of the slow-gain and fast-loss feature,
because these PDFs possess exponential tails, whence we infer the
characteristic loss and gain times $\tloss$ and $\tgain$, respectively. We
obtain $\tloss < \tgain$, for all the cases we have considered. It is
well-known that, in 3D turbulent flows, every point in the flow can be
classified into two topological classes~\cite{perry1987description,
cho+per+can90}: vortical regions or saddles, which are strain-dominated,
depending on whether the discriminant of the velocity-gradient-matrix is
positive or negative.  By using this discriminant, we show that the slow-gain
in energy of the particles is equally likely in vortical or strain-dominated
regions of the flow; in contrast, the fast-loss of energy occurs with greater
probability in the latter than in the former. 

%
%
%
 
The remainder of this paper is organized as follows. In
Section~\ref{chap4:mdls}, we introduce the models we use and the numerical
methods we employ to study them. Section~\ref{chap4:results} is devoted to a
presentation of our results. We discuss our results in the concluding
Sec.~\ref{chap4:con}.
  
\section{Model and numerical methods}
\label{chap4:mdls}

If the flow velocity at the position of the particle is $\uu$, then the motion
of a heavy particle is governed by the following equations:
\begin{subequations}
\begin{align}
\Xdot &= \vv \/, \label{eq:dxdt}\\
\vdot &= \frac{1}{\taup}\left[ \uu(\XX) - \vv \right] \/.
\label{eq:dvdt}
\end{align}
\label{eq:HIP}
\end{subequations}
here $\vv(t)$ and $\XX(t)$ denote, respectively, the velocity and position of
the particle at time $t$, and $\taup = (2 \mathit{a}^2\rhop)/(9\nu \rhof)$ is
the Stokes or response time of the particle, with $a$ and $\rhop$ the radius
and material density of the particle, respectively.  Equation~(\ref{eq:HIP})
is valid if (a) the radius of the
particle $\mathit{a} \ll \eta$, with $\eta$ the Kolmogorov dissipation scale of
the advecting fluid (or the particle-scale Reynolds number is very small),
(b) interactions between particles are negligible, (e.g., at low number 
densities of particles), (c) the particle density $\rhop \gg \rhof$, the fluid
density, (d) typical particle accelerations are much larger than the
acceleration because of gravity, and (e) the fluid velocity is not
affected by the particles. 

\subsection{Three-dimensional Navier-Stokes turbulence} 
\label{subsec:3d_turb}

We consider the motion of the particles described by Eqs. (\ref{eq:HIP}) in 3D,
homogeneous, and isotropic turbulent flows.  The velocity field $\uu(\xx,t)$ is
obtained by solving the three-dimensional (3D), incompressible, Navier-Stokes
equation, i.e.,
\begin{subequations}
\begin{align}
\partial_t\uu + \uu\cdot \grad\uu &= \nu \lap\uu - \grad p +\ff\/,\label{eq:mom}\\
\grad\cdot\uu &= 0\/, \label{eq:incom}
\end{align} 
\label{eq:ns}
\end{subequations}
where $p$, $\ff$, and  $\nu$ are the pressure, external
force, and the kinematic viscosity, respectively. 
To solve \Eq{eq:ns}) numerically, we use a pseudo-spectral method~\cite{Can88} 
with periodic boundaries and the $2/3$ de-aliasing rule.
Table~\ref{table:para} gives the parameters for our DNSs of the 3D 
Navier-Stokes equation~\footnote{Note that the value of $Re_\lambda$ given 
in Table \ref{table:para} is smaller than reported in other studies in the 
literature for the same resolution, e.g., 
Ref~\cite{yoshimoto2007self,bec2006acceleration}; in these 
studies, their expression for $\lambda$ contains an extra factor of $\sqrt{5}$
relative to our definition.} The Stokes number that
we use is $\St = \taup/\teta$. 
\begin{table*}
\bc
\caption{Parameters for our 3D runs ${\bf R1}$ and ${\bf R2}$ with
$N^3$ collection points, $\nu$ the coefficient of kinematic viscosity, $\delta
t$ the time step,  $N_p$ the number of particles, $k_{max}$ the largest
wave number in the simulation, $\eta$ and $\tau_\eta$ the dissipation
length and time scales, respectively, $\lambda$ the Taylor micro-scale,
$Re_\lambda$ the Taylor-micro-scale Reynolds number, $I_l$ the integral
length scale, and $T_{eddy}$ the large-eddy turnover time.}
\resizebox{1.0\linewidth}{!}{
\begin{tabular}{c c c c c c c c c c c c c}
\hline
Run & $N$ &	$\nu$	& $\delta t$ & $N_p$ & $Re_\lambda$ & $\kmax\eta$ & $\epsilon$ &
$\eta$ & $\lambda$ & $I_l$ & $\tau_\eta$ & $T_{\rm eddy}$ \\
\hline\hline
{\bf R1} & $256$ & $3.8\times 10^{-3}$ & $5\times10^{-4}$ & $40,000$ & $43$ & $1.56$ & $0.49$ &
$1.82\times10^{-2}$ & $0.16$ & $0.51$ & $8.76\times10^{-2}$ & $0.49$ \\

{\bf R2} & $512$ & $1.2\times 10^{-3}$ & $2\times10^{-4}$ & $100,000$ & $79$ & $1.21$ & $0.69$ &
$7.1\times10^{-3}$  & $0.08$ & $0.47$ & $4.18\times10^{-2}$ & $0.41$ \\
\hline
\end{tabular}}
\label{table:para}
\ec
\end{table*}
\section{Results}
\label{chap4:results}

We first allow the flow to develop until it reaches a statistically stationary
turbulent state; and then we introduce the particles. We also ignore the
transients until the heavy particles reach a nonequilibrium statistically
stationary state, which we monitor via the temporal evolution of
the total energy of the particles. In this nonequilibrium state, the PDF of 
any component $v_k$ of the velocity, of a heavy particle, 
is a Gaussian with zero mean and a variance 
\begin{equation}
\bra{v^2} \approx \frac{\urms^2}{1+\StT}\/,
\end{equation}
where $\StT$ is the Stokes number defined with respect to the
large-eddy-turnover time.  The auto-correlation function $C(t) \equiv
\bra{v_k(0)v_k(t)}/\bra{v_k^2}$, at large $t$, decays with a time scale that is
shorter than the large-eddy-turnover time of the flow (see
Appendix~\ref{appendix} for details).

As we have mentioned above, we follow the Lagrangian-tracer studies of
Refs.~\cite{xu2014flight,xu2013power}, and we characterize the irreversibility
of the dynamical system formed by the particles by calculating the statistics
of the energy increments $W$ and the power $p$: 
\begin{subequations}
\begin{align}
W(\tau) &\equiv E(t+\tau) - E(t)  \/, \\
p &\equiv \vv\cdot \frac{d\vv}{dt} \/,
\label{eq:power}
\end{align}
\end{subequations}
where $E \equiv (1/2)\mid\vv\mid^2$ is the energy-per-unit-mass.  

\subsection{Statistics of energy increments}
\label{W}
\begin{figure*}
\begin{center}
\includegraphics[width=0.32\linewidth]{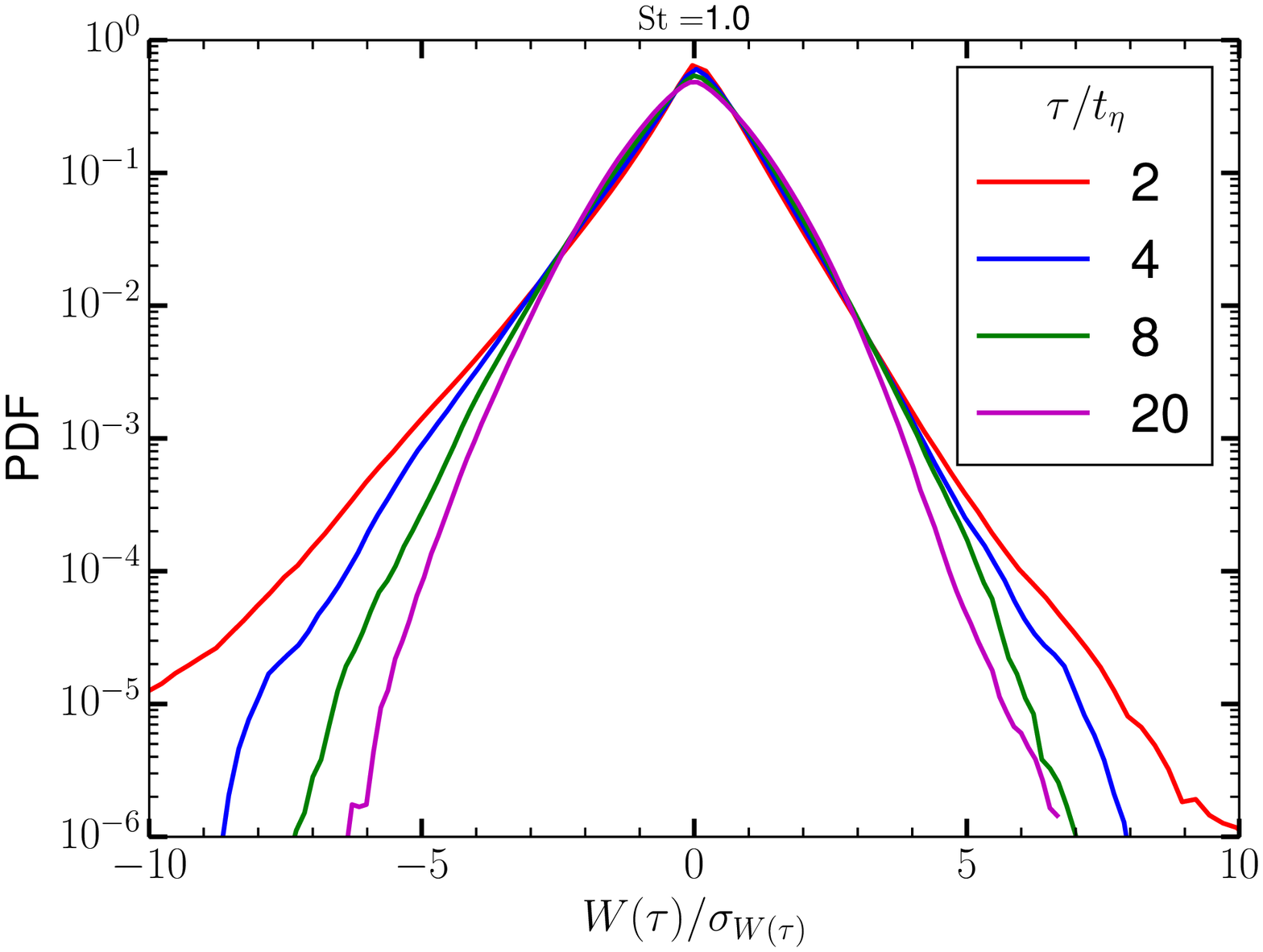}  
\put(-130,100){\bf (A)}
\includegraphics[width=0.32\linewidth]{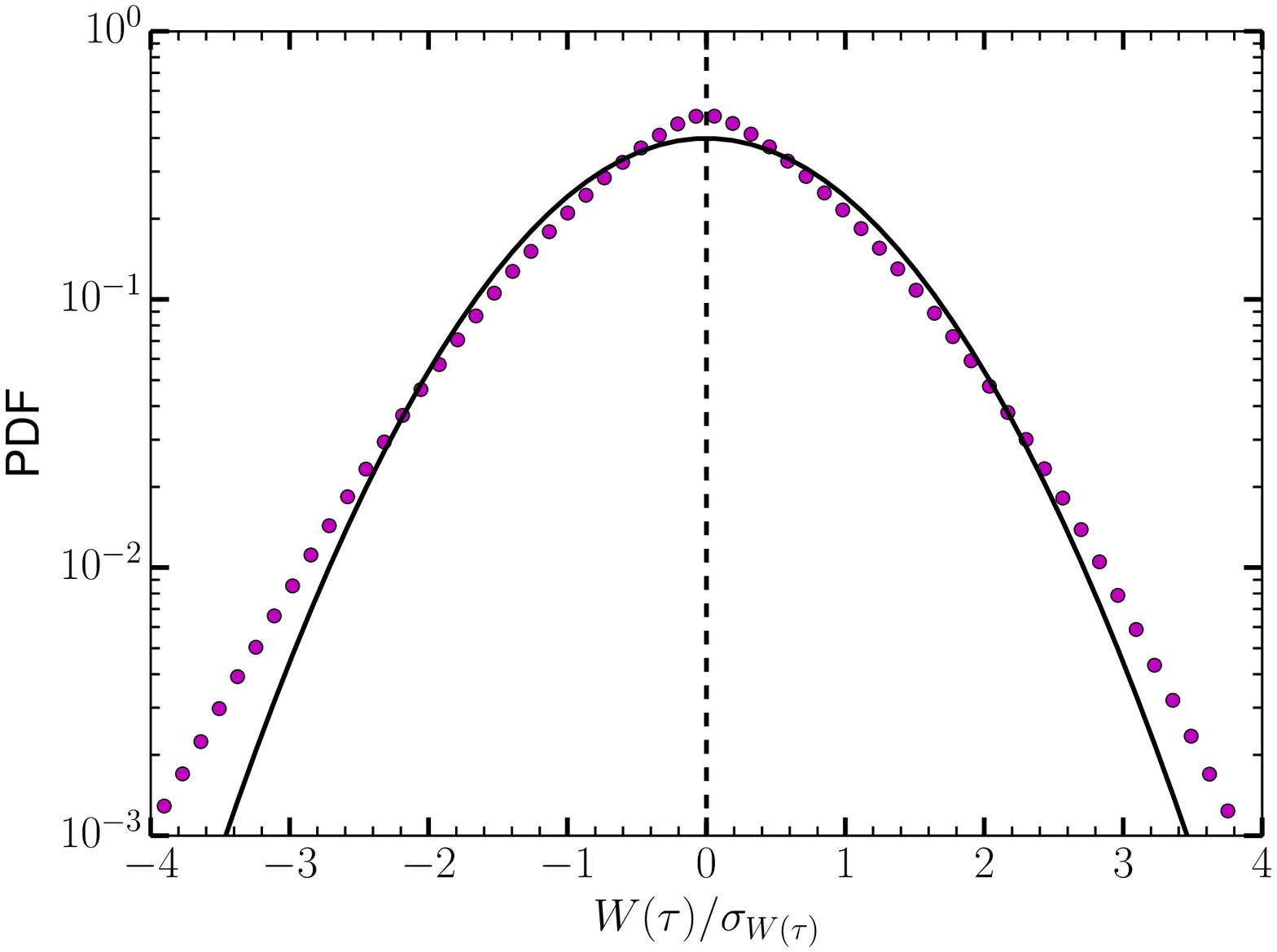}  
\put(-130,100){\bf (B)}
\includegraphics[width=0.32\linewidth]{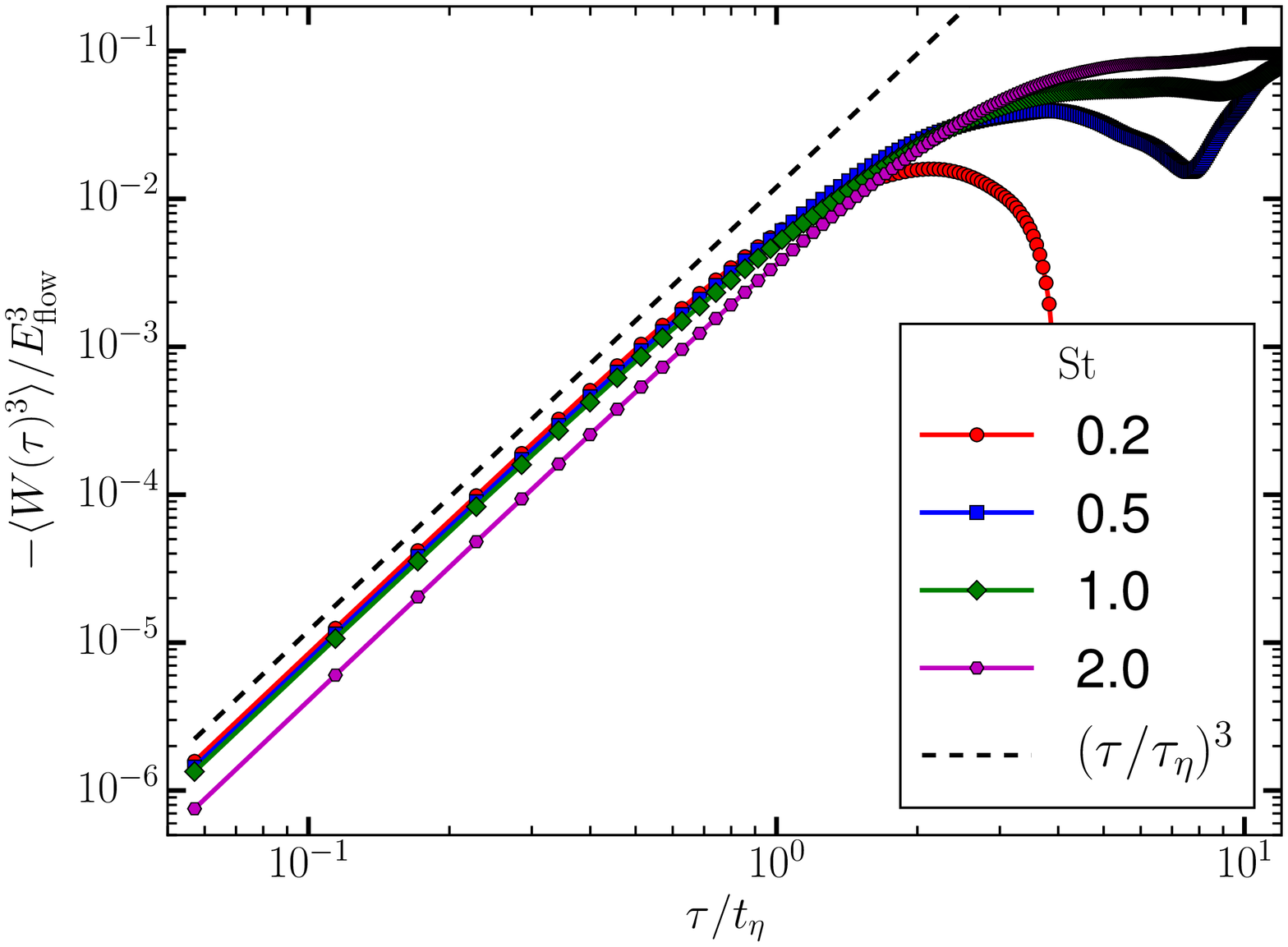} 
\put(-130,100){\bf (C)}
\caption{(Color online) (A) Probability density functions of the energy increments,
$W(\tau)$ for the different values of the time lags, $\tau$, for $\St=1$.  (B)
Probability density function of $W(\tau)$ for $\tau = 20\teta$ and $\St=1$ (magenta circles), 
compared with a normal distribution with zero mean and unit variance (solid black line).
(C) The third moment of the PDF of $W(\tau)$ as function of $\tau$,
for different values of $\St$.} 
\label{fig:pdfW}
\end{center}
\end{figure*}

In \subfig{fig:pdfW}{A}, we plot the PDF of the energy increment, $W(\tau)$,
across a time-scale $\tau$ (normalized by the dissipation time $\teta$), for
several different values of $\tau$ and $\St = 1$. A careful look at this
figure shows that this PDF is asymmetric about zero, with an asymmetry that is
most pronounced for small $\tau$.  Even for large $\tau$, these PDFs do not
approach a Gaussian distribution, as we demonstrate in \subfig{fig:pdfW}{B}.
In \subfig{fig:pdfW}{C} we plot the simplest characterization of the asymmetry
of the PDF of $W(\tau)$, namely, its third moment $\bra{W^3(\tau)/\Eflow^3}$,
as a function of $\tau$, for different values of $\St$, where the
characteristic energy of the flow 
$\Eflow\equiv (1/2)\langle \uu^2 \rangle$ 
is used to non-dimensionalize  $W$.  As we  
expect~\cite{xu2014flight}, at small $\tau$, the third-moment scales as
$\tau^3$, because $W(\tau)$ is smooth, so it can be Taylor expanded 
at small $\tau$. 

\subsection{Statistics of the power input}
\label{power}

\begin{figure}
\begin{center}
\includegraphics[width=0.9\linewidth]{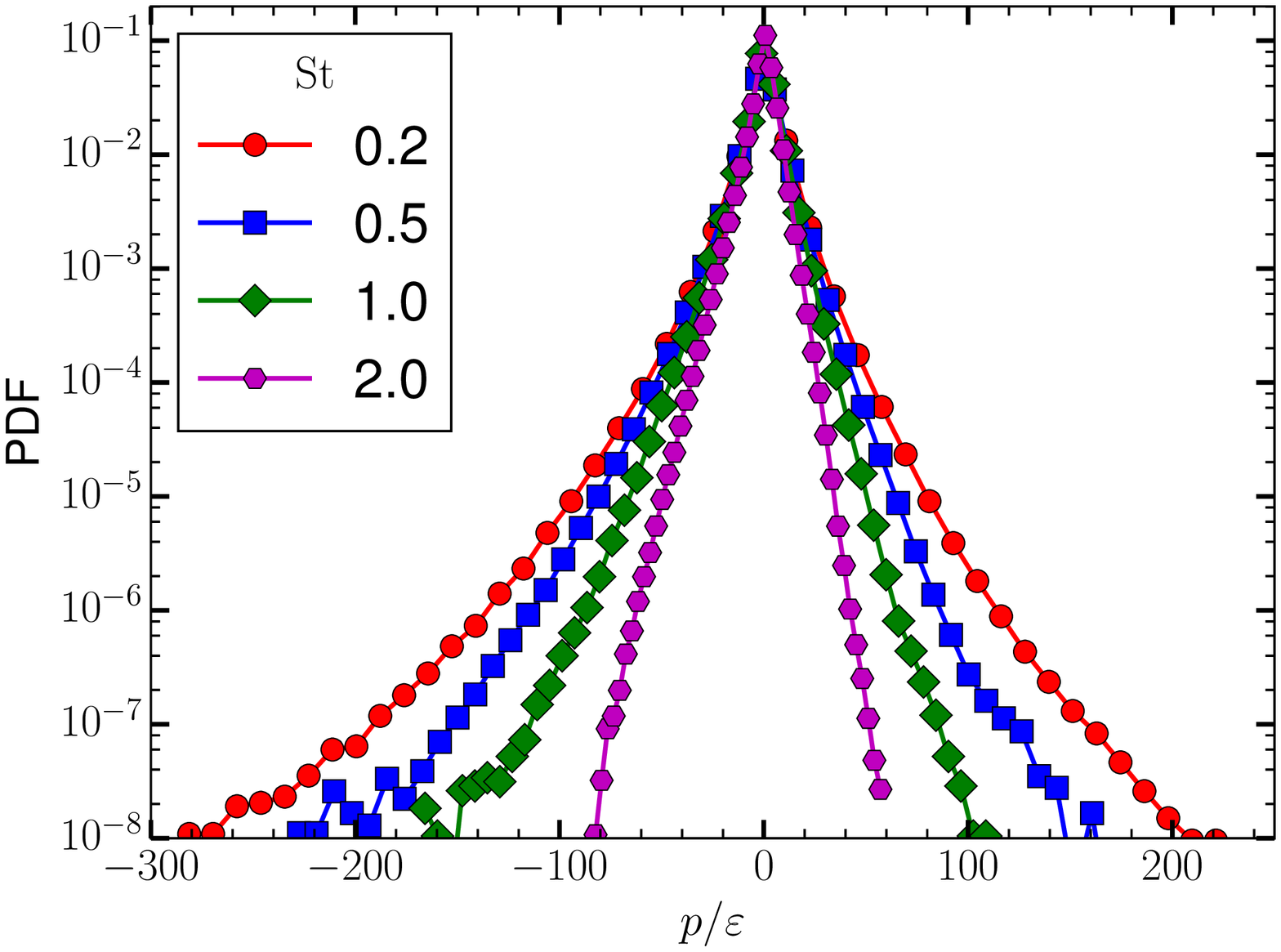}
\put(-30,140){\bf (A)}\\
\includegraphics[width=0.9\linewidth]{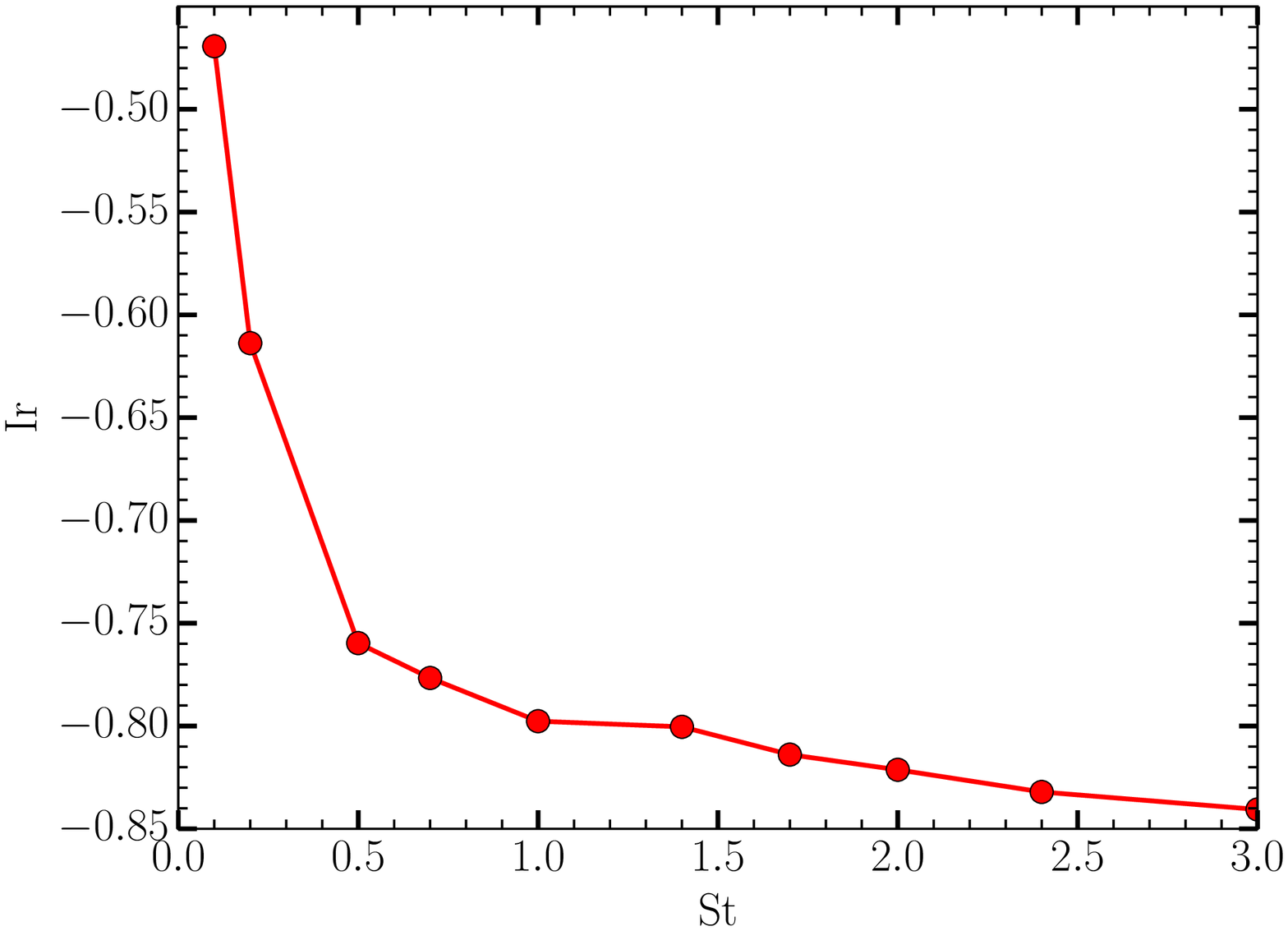}
\put(-30,140){\bf (B)}
\caption{(Color online) (A) The PDF of the non-dimensionalized
  power input $p/\varepsilon$ to the particle by flow. (B) The measure of time
irreversibility $\Ir$, defined in \Eq{eq:Ir}, as a function of $\St$.
} 
\label{fig:pdf_power}
\end{center}
\end{figure}

We now plot in \subfig{fig:pdf_power}{A}, the PDF of the power-input $p$ to the
particle per-unit-mass; $p$ is normalized by $\eps$, the
rate-of-energy-dissipation of the flow . A careful look at the figure shows
that the tails of the PDF are negatively skewed; they fall off more slowly on
the negative side than on the positive side.  This can be quantified by
plotting the skewness of these PDFs, which, following
Ref.~\cite{xu2014flight}, we define as the irreversibility parameter:
\begin{equation}
\Ir = \frac{\bra{ p^3 }}{\bra{p^2}^{3/2}} \/.
\label{eq:Ir}
\end{equation} 
In \Fig{fig:pdf_power} we plot $\Ir$ as a function of $\St$.  As $\St\to0$ we
expect that $\Ir$ should approach its value for Lagrangian tracers. We find
that $\Ir$ remains negative for all $\St$; in particular, its magnitude
increases sharply, at small $\St$, but this increase slows down at about $\St
\simeq 1$.  

\subsection{Time scales of the gain and loss of energy}

We now provide a direct and natural quantification of the slow-gain and
fast-loss phenomenon by analyzing the time series of $p$ as follows:  Let
$\tplus$ ($\tminus$) be the time over which $p$ has a positive (negative) sign,
i.e., the particle gains (loses) energy. These times are the first-passage
times, from positive to negative values or vice versa, of the random variable
$p$.  The PDFs of such first-passage times are called persistence PDFs; if a
persistence PDF has a power-law tail the exponent of the power-law is called
the persistence exponent~\cite[see, e.g., ][for the use of persistence in
various problems of nonequilibrium statistical mechanics]{maj99,bra+maj+sch13}.
The same idea has been used to calculate the persistence PDFs of residence
times of tracers~\cite{per+ray+mit+pan11} and heavy inertial particles in
topological structures in two dimensional~\cite{gupta2014statistical} and
3D~\cite{bhatnagar2016deviation} turbulent flows. 

\begin{figure}
\begin{center}
\includegraphics[width=0.90\linewidth]{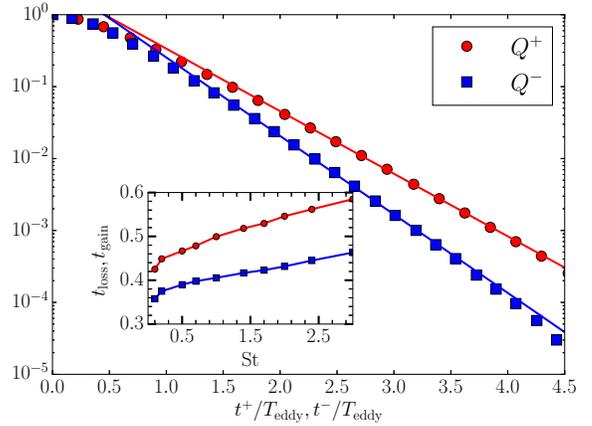}
\caption{(Color online) The two cumulative PDFs, $\Qplus$ and $\Qminus$, of the
times for which the $p$ remains respectively positive (red) and negative
(blue), for $\St=1$.  The two straight lines are linear fits to the tail of the data.  The
slope of these straight lines are $\tgain$ and $\tloss$. These are plotted as a
function of $\St$ in the inset, $\tgain$ (red circles) and $\tloss$ (blue
squares). These time scales are scaled by large eddy turn over time 
of the flow $\Teddy$.}
\label{fig:pers_p}
\end{center}
\end{figure}

From the time-series of $p$ we calculate the cumulative probability
distribution (CDF) of both $\tplus$ and $\tminus$, which we denote by $\Qplus$
and $\Qminus$, respectively~\footnote{ It is generally difficult to obtain
reliable information about the tail of a PDF by plotting histograms because of
possible binning errors. Hence, instead of studying the tail of the PDF, we
have calculated the CDF of the power of the particles by using the rank-order
method~\cite{mit+bec+pan+fri05}, which leads to the CDF that is free from
binning errors.  By definition, the CDF of a random variable $s$ is given by,
\unexpanded{$Q(s) \equiv \int^{s}_{0}P(s)ds$}, where $P$ and $Q$ are,
respectively, the probability density function and the cumulative distribution
function.  To calculate the CDF of a set of data, with $N$ samples, by using
the rank-order method, we sort the data in \textit{decreasing} order, assign
the maximum value rank $1$, the next value the rank $2$, and so on. The
quantity we plot in the vertical axis of \Fig{fig:pers_p} is this rank divided
by the sample size $N$; clearly, this is equal to $1-Q(s)$. This method is best
suited for studying tails of CDFs. If we are interested in the behavior of the
CDF for small values of its arguments, it is necessary to sort the data in
\textit{increasing} order, and then follow the method given above.}  
These two
CDFs, for $\St=1$, are plotted in \Fig{fig:pers_p} on log-lin scales. Clearly
both $\Qplus$ and $\Qminus$ have exponential tails, with characteristic time
scales  $\tgain$ and $\tloss$, respectively.  This implies that the
corresponding PDFs also possess exponential tails, with the same characteristic
time scales.  These two time scales are plotted, as functions of $\St$, in the
inset of \Fig{fig:pers_p}, from which we infer that, for all $\St$, $\tgain <
\tloss$, which is a natural quantification of the slow-gain and fast-loss
feature.

\subsection{Irreversibility and the topology of the flow}

\begin{figure}
\begin{center}
\includegraphics[width=0.9\linewidth]{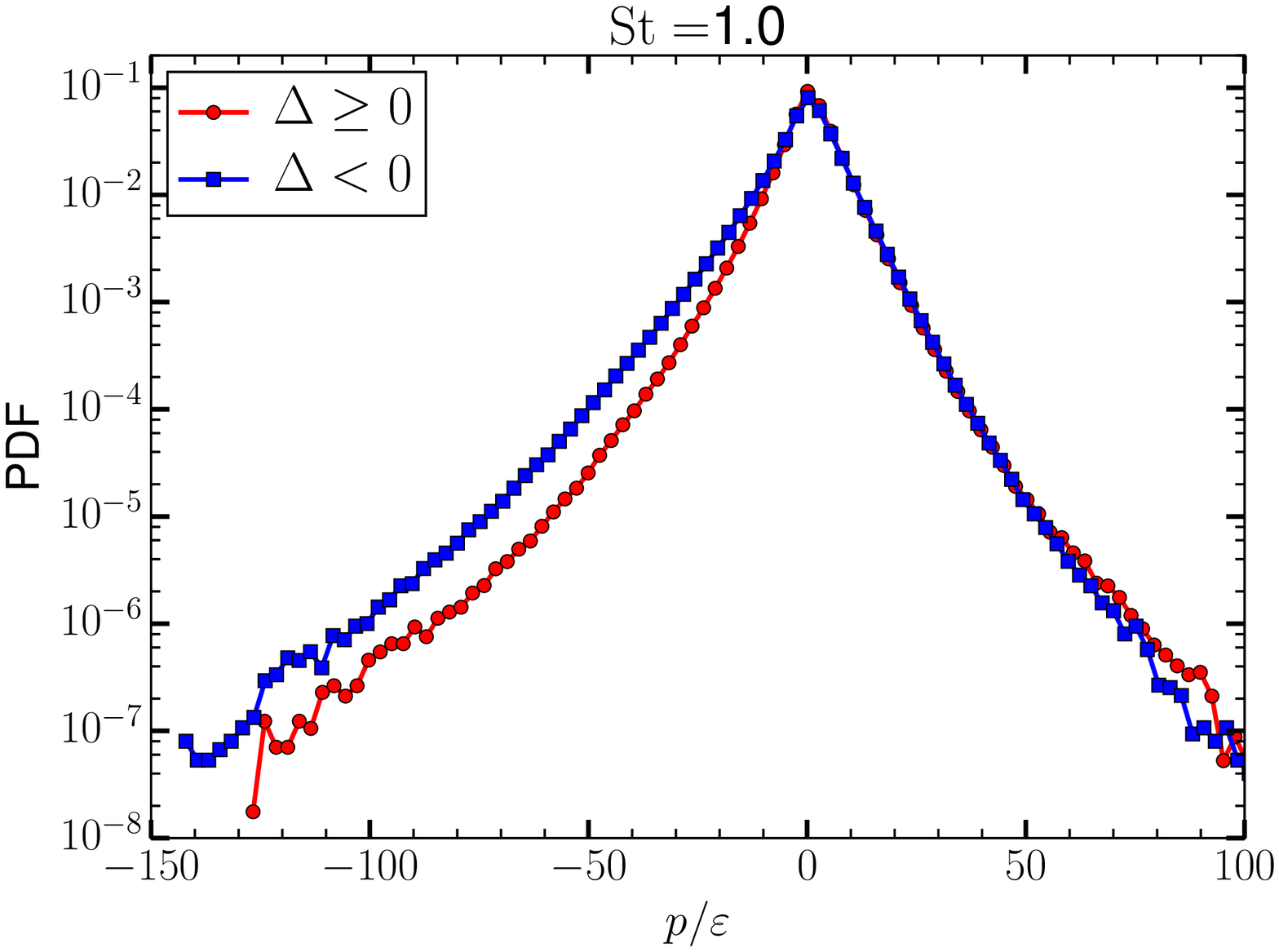} \\
\includegraphics[width=0.9\linewidth]{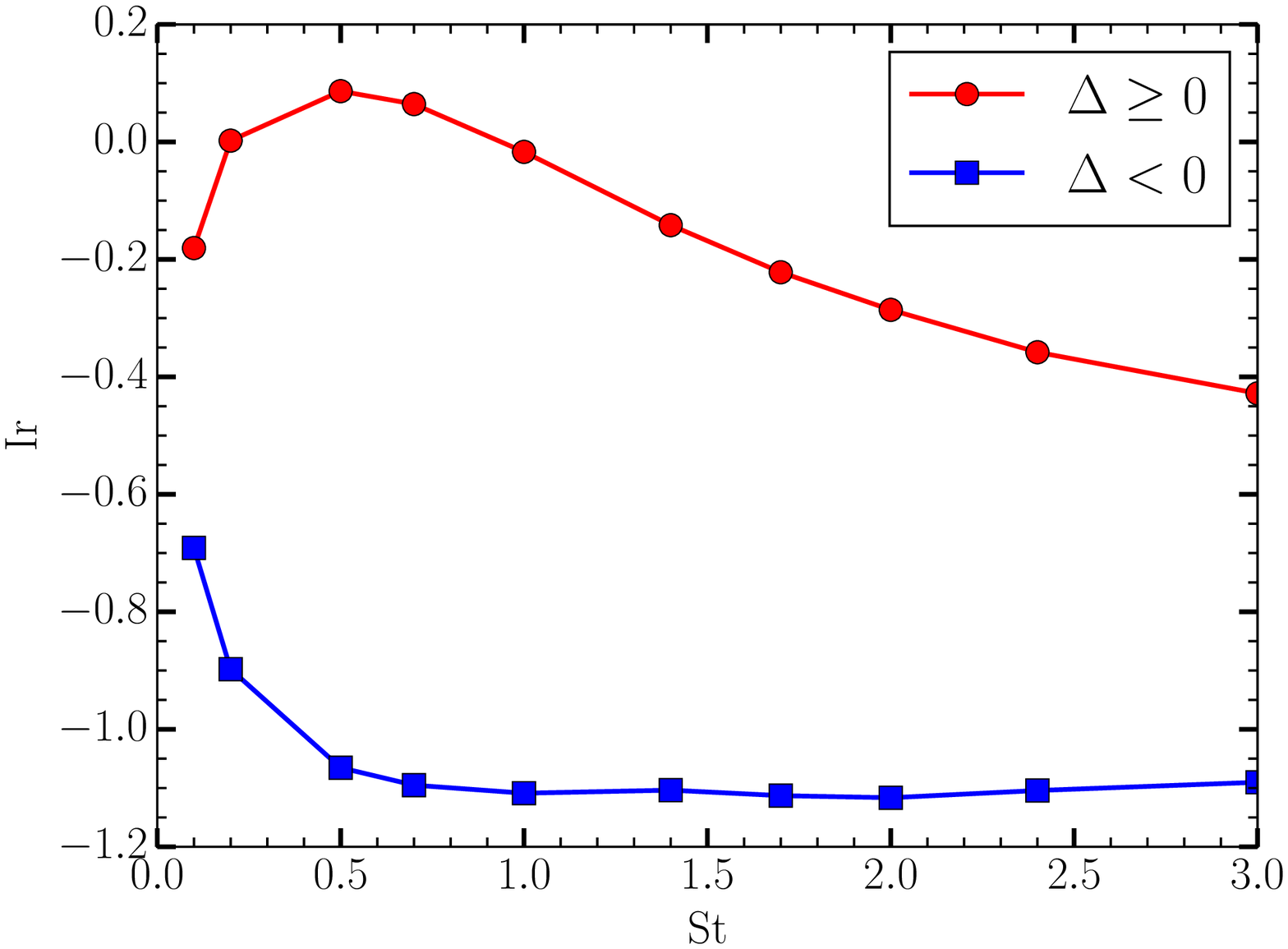}
\caption{(Color online) (top panel) The PDFs of the non-dimensionalized
  power input $p/\eps$ obtained separately from  vortical (red) and 
strain-dominated (blue) regions of the flow, for $\St=1$.  
(bottom panel) Contributions to the irreversibility parameter $\Ir$
from  vortical (red) and strain-dominated (blue) regions of the flow.} 
\label{fig:powercDe}
\end{center}
\end{figure}

The topology of a 3D vector field can be characterized by its gradient-matrix.
A $3\times 3$ matrix, $\BB$ has three invariants, namely, its trace $\Tr\BB =
\lambda_1+\lambda_2+\lambda_3$, $Q \equiv \lambda_1\lambda_2
+\lambda_2\lambda_3 +\lambda_3\lambda_1$, and its determinant $\Det\BB =
\lambda_1\lambda_2\lambda_3$, where $\lambda_1$, $\lambda_2$, and $\lambda_3$
the are three eigenvalues of $\BB$. If the vector field is incompressible, like
our flow velocity field, there are only two invariants, because the trace of
the velocity-gradient matrix is zero everywhere. We consider incompressible
turbulent flows, so the velocity-gradient matrix is a random matrix with zero
trace; it is conventional~\cite{cho+per+can90} to denote its two invariants by
the symbols $Q$ and $R \equiv -\Det \BB$. Depending on the values of $Q$ and
$R$, four different types of flow topologies are possible: two are elliptic (or
vortical) points, with a third stable/unstable direction, and two are saddles,
with axial or bi-axial strain. Whether the flow at a point is a topological
vortex or a saddle depends on the sign of the discriminant, 
$\Delta \equiv (27/4)R^2+Q^3$, of the
characteristic equation of the velocity-gradient matrix; it is positive in
vortical regions and negative in strain-dominated saddles. We have argued above
that the particles lose energy to fast, small-length-scale eddies and gain
energy from large-length-scale eddies. The topological structures are
small-length-scale properties; hence, by the usual assumption of
length-scale-separation in turbulence, we expect that the gain in energy, which
occurs in large-scale eddies, does not depend on the topology of the flow. By
contrast, the loss in energy occurs in small-length-scale eddies, which are
intimately connected with the topological structures we have described above.
It has been established recently that heavy particles, in 3D turbulent flows,
spend more time in strain-dominated regions than in vortical
regions~\cite{bhatnagar2016deviation}; consequently, we expect that the loss of
energy occurs more in strain-dominated regions than in vortical regions in the
flow. To check the validity of this expectation, we plot, in the top panel of
\Fig{fig:powercDe}, the PDFs of $p$, obtained separately from regions with
saddles and vortices. There is no distinction between these two PDFs for
positive $p$, i.e., when the particles gain energy. By contrast, when $p$ is
negative, i.e., when the particles lose energy, this loss is more likely to
occur in strain-dominated flow regions than in vortical ones.  This is also
confirmed in the bottom panel of \Fig{fig:powercDe}, where we plot the
contribution to the irreversibility parameter $\Ir$, obtained separately from
vortical and strain-dominated regions, for several different values of $\St$;
in particular, the contribution from the vortices is significantly smaller than
that from the saddles, which shows that the dominant contribution to the
skewness of the PDF of the power comes from the saddles.    

\section{Conclusions}
\label{chap4:con}

We have carried out a detailed numerical study of the time irreversibility of
the dynamics of heavy particles in 3D, statistically homogeneous and isotropic
turbulent flows. We have shown that these particles, which follow \Eq{eq:HIP},
reach nonequilibrium statistically stationary states. We have characterize
these states by calculating variety of PDFs and auto-correlation functions. The
simplest of these are PDFs and  auto-correlation functions of the velocity
components; we have shown that these PDFs are close to Gaussian.  We have also
computed the PDFs of the increments of the particle's energy $W(\tau)$, for
different values of $\tau$, and shown that these PDFs are non-Gaussian and
skewed towards negative values. This implies that, on average, particles gain
energy over a period of time that is longer than the duration over which they
lose energy. For passive Lagrangian tracers, this phenomenon, has been called a
flight-crash effect in Ref.~\cite{xu2014flight};  we simply refer to it as slow
gain and fast loss. We have also found that the third moment of $W(\tau)$
scales as $\sim \tau^3$, at small values of $\tau$.  

We have computed the PDFs of the scaled power input $p$, for different values
of $\St$, and shown that it is negatively skewed. This negative skewness
provides us a measure of the time irreversibility $\Ir$. We have demonstrated
that the magnitude of $\Ir$ increases with $\St$, sharply for small $\St$, but
more slowly thereafter (at about $\St \approx 1$).  These qualitative features
can also be captured by models in which the flow velocity is obtained from
stochastic models with non-zero correlation time~\cite{aks_thesis}. 
 
Our study has led to a direct and natural measure of the slow-gain and
fast-loss of energy. Specifically, we have calculated the PDFs of $\tplus$ and
$\tminus$, the times over which $p$ has, respectively, positive or negative
signs.  These PDFs have exponential tails, from which we have inferred the
characteristic loss and gain times $\tloss$ and $\tgain$, respectively.  We
have shown $\tloss < \tgain$, for all the values of $\St$ we have considered.
Furthermore, we have shown that the slow-gain in energy of the particles is
equally likely in vortical or strain-dominated regions of the flow; in
contrast, the fast-loss of energy occurs with greater probability in the latter
than in the former.

Time irreversibility for Lagrangian tracers, advected by turbulent flows,
arises solely because of the time-irreversible nature of such flows. In
contrast, for the case of heavy particles, time irreversibility arises because
of two reasons: (a) turbulent flows, which advect such particles, are
irreversible; and (b) the Stokes drag, exerted by the flow on the particle, is
dissipative.  The separation of the effects of particle inertia and turbulence
on time irreversibility is non-trivial. Our study has shown how the effect of
inertia can be captured clearly by the dependence of $\Ir$ on $\St$, which we
have shown in Fig.~\ref{fig:pdf_power}.

The time irreversibility for Lagrangian tracers, advected by turbulent flows,
has been studied theoretically, numerically, and experimentally (see, e.g.,
Refs.~\cite{xu2013power,xu2014flight}). Our study has carried out analogous
theoretical and numerical studies for heavy particles advected by turbulent
flows; and we have obtained clear signatures for such irreversibility, which
can be measured in heavy-particle-laden flows. We hope, therefore, that our
studies will stimulate experimental investigations of time irreversibility in
such  heavy-particle-laden flows.

\section{Acknowledgments}

This work has been supported  in part by the Swedish Research Council under
grants 2011-542 and 638-2013-9243 (DM), Knut and Alice Wallenberg Foundation
(DM and AB) under the project ”Bottlenecks for particle growth in turbulent
aerosols” (Dnr.  KAW 2014.0048), Council of Scientific and Industrial
Research (CSIR), University Grants Commission (UGC), and Department of Science
and Technogy (DST India) (AB and RP).  We thank SERC (IISc) for providing
computational resources. DM thanks the Indian Institute of Science for
hospitality during the time some of these calculations were initiated.  RP
thanks NORDITA for hospitality during the period in which this paper was being
written. We thank Prasad Perlekar and Samriddhi Shankar Ray for useful discussions.


\appendix
\subsection{Characterization of the statistically stationary turbulent state}
\label{appendix}
\begin{figure*}[hb] 
\begin{center}
\includegraphics[width=0.32\linewidth]{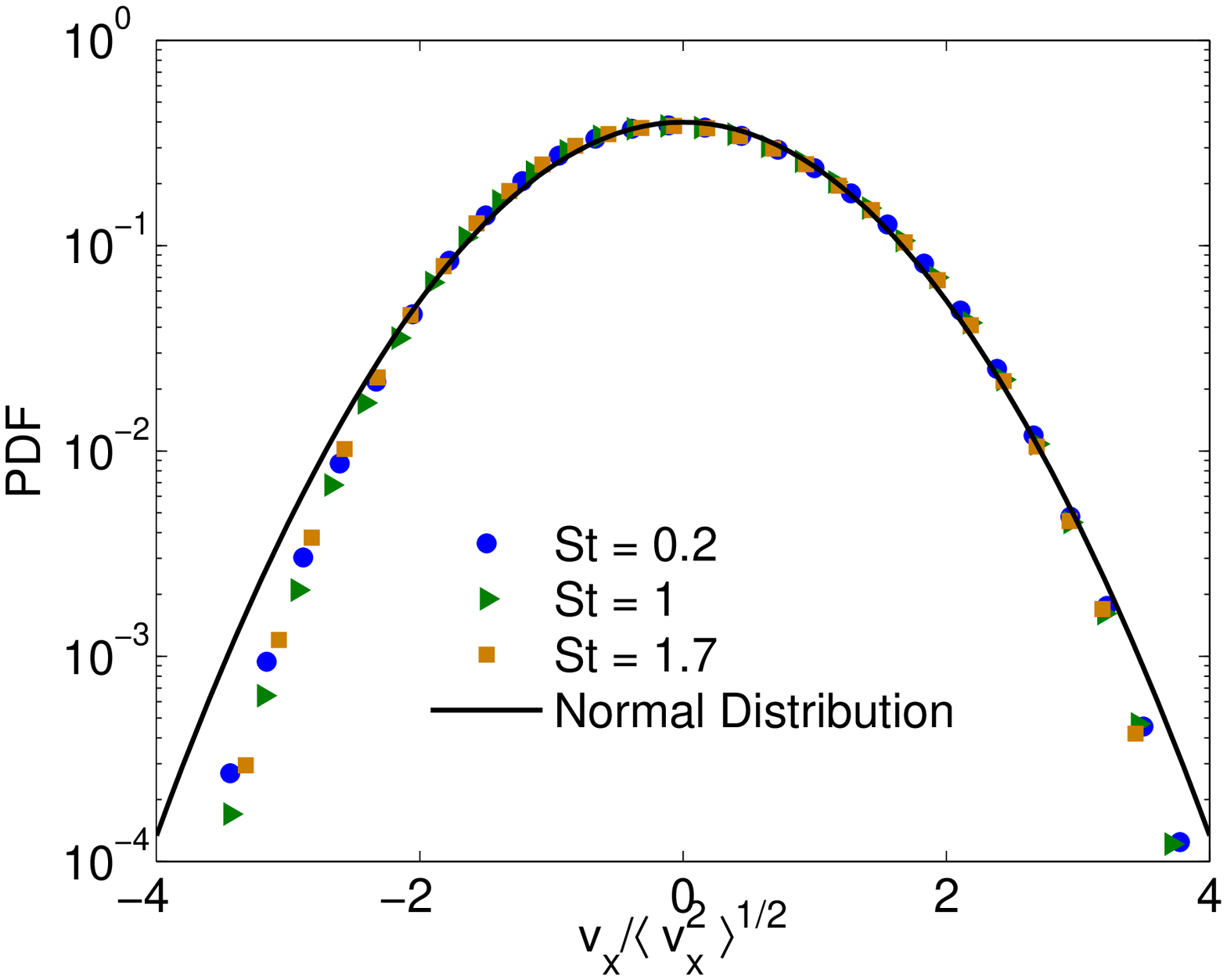}
\includegraphics[width=0.32\linewidth]{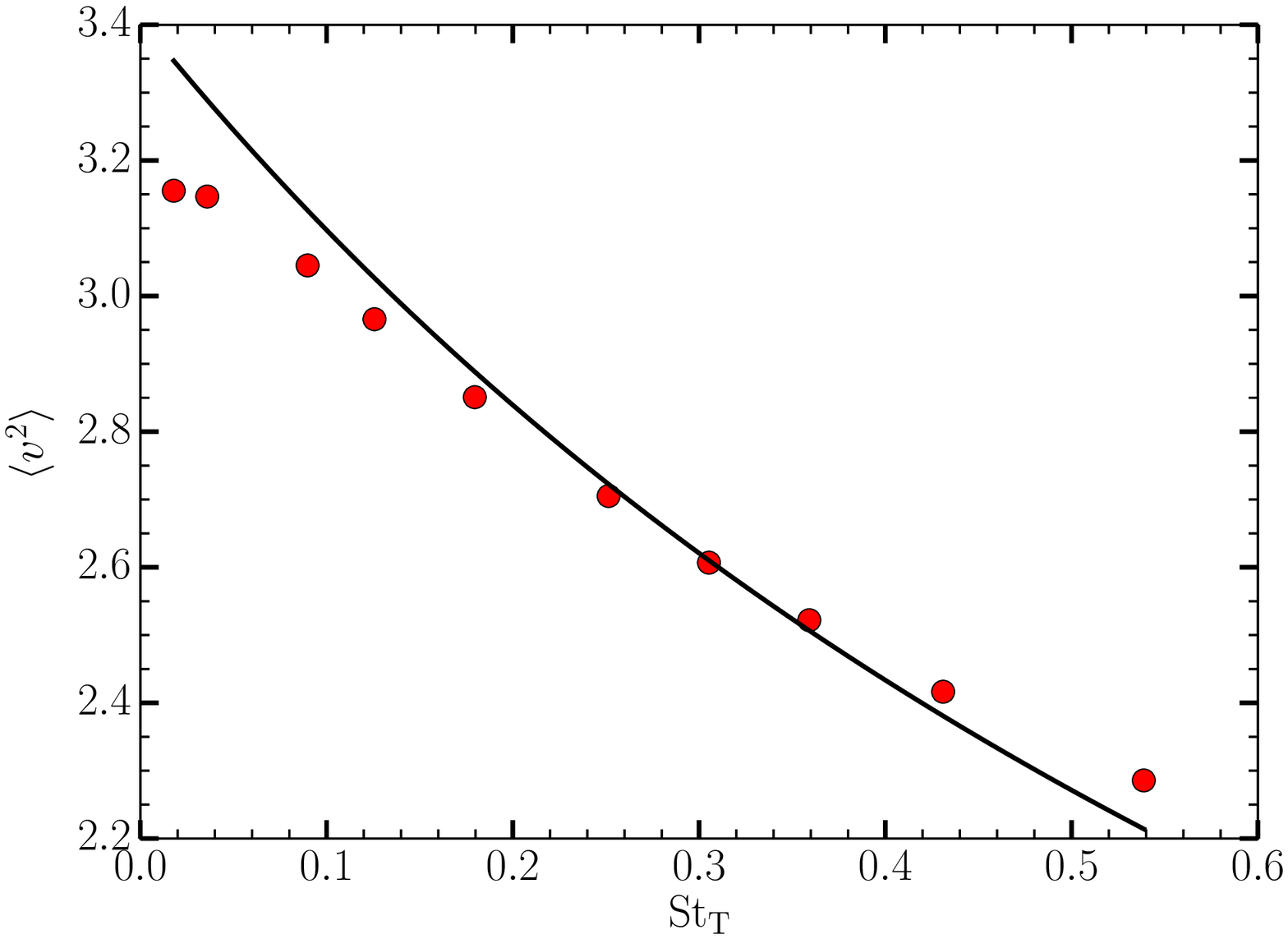} 
\includegraphics[width=0.32\linewidth]{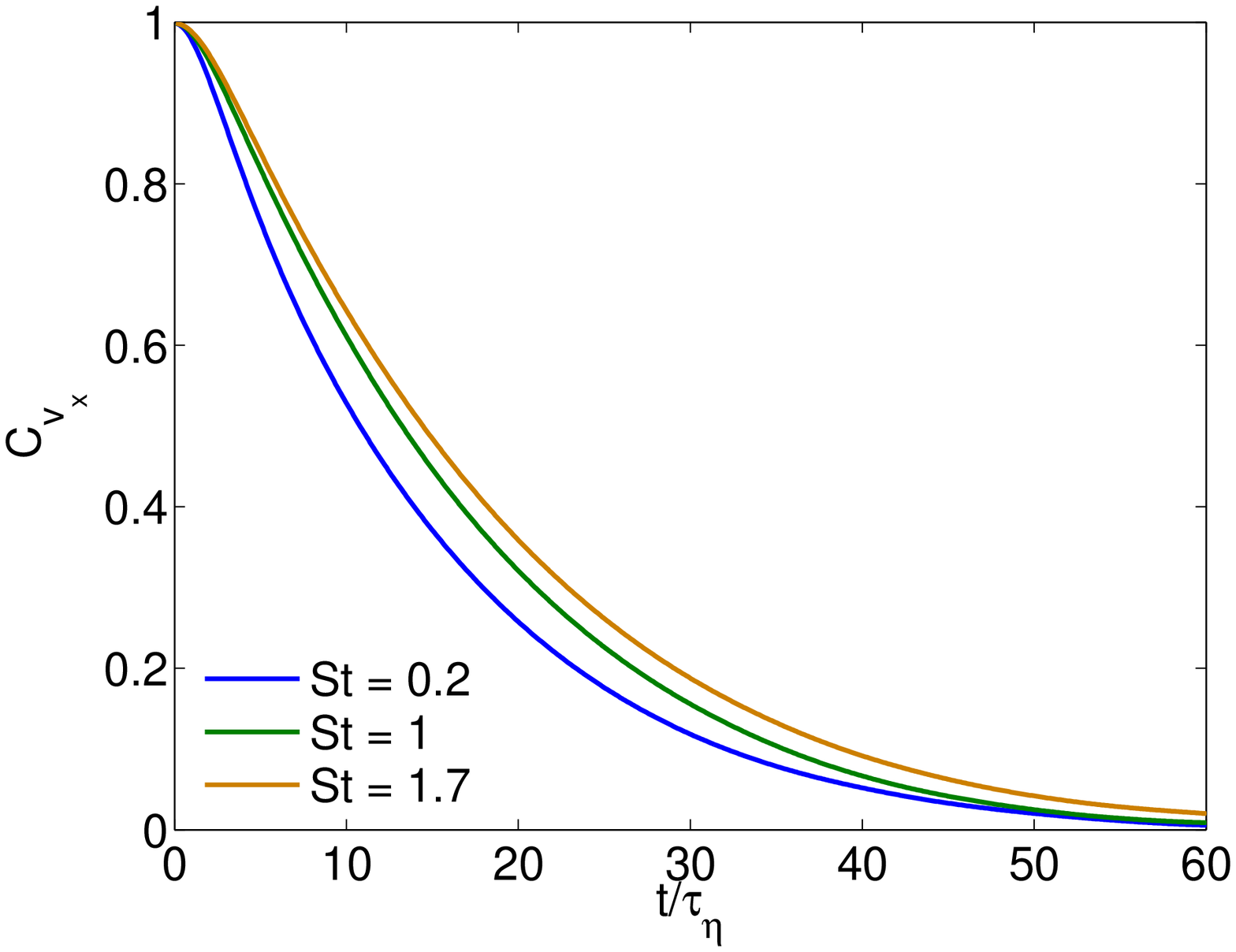}
\caption{(Color online)  (left panel) The PDFs of the $x$ component of the velocity of the particle.
The black solid line shows a normal distribution with mean zero and 
standard deviation unity.
(middle panel) The mean of $v^2$ plotted as a function of the Stokes-number 
defined, by $\Teddy$, as $\StT=\taup/\Teddy$; the black solid line shows 
the plot of $\langle u^2 \rangle/(1+\StT)$ as a function of $\StT$.
(right panel) The auto-correlation function $C_{v_x}$ of the $x$ component 
of the velocity of the particle.}
\label{fig:evst}
\end{center}
\end{figure*}
In Fig.~\ref{fig:evst} we show plots, of PDFs of $x$ component $v_x$ of the 
velocity of the particle (left panel). These PDFs are close to a 
Gaussian distribution. (middle panel) Shows the mean of $v^2$ plotted as a 
function of the Stokes-number 
defined, by $\Teddy$, as $\StT=\taup/\Teddy$; the black solid line shows 
the plot of $\langle u^2 \rangle/(1+\StT)$ as a function of $\StT$.   
(right panel) of \Fig{fig:evst} shows the auto-correlation function 
\begin{equation}
C(t) \equiv \frac{\bra{ v_x(t) v_x (0)}}{\bra{v_x^2}}
\label{eq:autocor}
\end{equation}
of the $x$ component of $\vv$. The auto-correlation functions decay at 
large times. The characteristic decay time decreases with $\St$. 
\end{document}